\documentclass[submission,copyright,creativecommons]{eptcs}
\providecommand{\event}{TiCSA 2023}

\usepackage{underscore}         

\usepackage{xspace}
\usepackage{amssymb}
\usepackage[dvipsnames,table]{xcolor}
\usepackage{hyperref}
\usepackage[capitalise,nameinlink]{cleveref}
\usepackage{url}
\usepackage{booktabs}
\usepackage{relsize}

\usepackage[scaled=0.85]{beramono}
\usepackage[T1]{fontenc}
\usepackage[utf8]{inputenc}
\usepackage{multirow}

\usepackage[normalem]{ulem}
\usepackage[textsize=scriptsize]{todonotes}
\newcommand{\bgtext}[1]{%
  \bgroup\markoverwith {\textcolor{#1}{\rule[-0.5ex]{2pt}{11pt}}}\ULon}

\usepackage{tikz}
\usetikzlibrary{arrows,calc,shapes.arrows,shapes.multipart,shapes.geometric, positioning, angles, shadows, automata,fit,decorations.markings,backgrounds, shadows.blur}
\usepackage{forest} 
\usepackage{ifthen}
\forestset{
  declare count register=disjuncts from,
  disjuncts from'=1,
  declare count register=concrete from,
  concrete from'=1,
  concrete colour/.code={\colorlet{concretecol}{#1}},
  abstract colour/.code={\colorlet{abstractcol}{#1}},
  draw colour/.code={\colorlet{drawcol}{#1}},
  concrete colour=blue!85!cyan!40,
  abstract colour=blue!85!cyan!15,
  draw colour=darkgray,
  /tikz/mandatory/.style={circle, fill=drawcol, draw=drawcol,inner sep=2pt},
  /tikz/optional/.style={circle, draw=drawcol, fill=white,inner sep=2pt},
  /tikz/concrete/.style={fill=concretecol, draw=drawcol},
  /tikz/abstract/.style={fill=abstractcol, draw=drawcol},
  /tikz/or/.style={},
  mandatory/.style={edge label={node [mandatory] {}}},
  optional/.style={edge label={node [optional] {}}},
  or/.style={for first={disjunct={#1}}},
  disjunct/.style={
    tikz+={
      \path (.parent) coordinate (A) -- (!u.children) coordinate (B) -- (!ul.parent) coordinate (C) pic {angle};
      \foreach \i/\j in {#1} {
        \path (!u\i.parent) coordinate (A) -- (!u.children) coordinate (B) -- (!u\j.parent) coordinate (C) pic [fill=drawcol,angle radius=8pt] {angle};
      }
    }
  },
  alternative/.style={for first={alternatives={#1}}},
  alternatives/.style={
    tikz+={
      \path (.parent) coordinate (A) -- (!u.children) coordinate (B) -- (!ul.parent) coordinate (C) pic {angle};
      \foreach \i/\j in {#1} {
        \path (!u\i.parent) coordinate (A) -- (!u.children) coordinate (B) -- (!u\j.parent) coordinate (C) pic [fill=none,draw=drawcol,angle radius=8pt] {angle};
      }
    }
  },
  disjunction tree/.style={
    where={isodd(n_children())}{
      for n={int((n_children()+1)/2)}{calign with current},
    }{
      calign=midpoint,
    },
    before typesetting nodes={
      for nodewalk={
        filter/.wrap pgfmath arg=
          {{level>=##1}{n_children()>1}}{(disjuncts_from)}
      }{
        or
      },
    },
    for tree={
      parent anchor=children,
      child anchor=parent,
      l=-20mm,
      blur shadow,
      rounded corners,
      text height=1ex,
      text depth=0.2ex,
      font=\sffamily\scriptsize,
      where={level()>=concrete_from()}{
        concrete,
      }{
        abstract,
      },
    },
  },
}

\newcommand{\myparagraph}[1]{\medskip\noindent\textbf{#1} }

\newcommand{\wrap}[1]{\begin{tabular}{@{}c@{}}#1\end{tabular}}

\newcommand{\Uppex}{\textcolor{blue!55!green!85!black}{\textsf{Uppex}}\xspace}

\tikzstyle{arrow}=[-stealth,thick,rounded corners=2pt,font=\sffamily\scriptsize,
                   above,inner sep=1pt,align=center]
\tikzstyle{arrS}=[arrow,blue!80!black]
\tikzstyle{arrL}=[arrow,purple!70!black,densely dotted]
\tikzstyle{arrN}=[arrow,red!70!black,densely dashed,rounded corners=5pt]
\tikzstyle{arrD}=[arrow,green!60!black,double]
\newcommand{\arrow}[1][arrow]{\raisebox{2pt}{\tikz{\draw[#1](0,0)--(0.5,0);}}\xspace}

\usepackage{listings}
\title{Spreadsheet-based Configuration of Families of Real-Time Specifications} 
\newcommand{\titlerunning}{Spreadsheet-based configuration of Families of Real-Time Specifications}
\author{
    José Proença
        \institute{CISTER and University of Porto, Portugal}
        \email{jose.proenca@fc.up.pt}
    \and
    David Pereira \qquad
    Giann Spilere Nandi
        \institute{CISTER, Polytechnic Institute of Porto, Portugal}
        \email{\{drp,giann\}@isep.ipp.pt}
    \and
    Sina Borrami \qquad
    Jonas Melchert
        \institute{Alstom}
        \email{\{sina.borrami,jonas.melchert\}@alstomgroup.com}
}
\newcommand{\authorrunning}{J. Proença, D. Pereira, G.S. Nandi, S. Borrami, and J. Melchert}

\hypersetup{
  bookmarksnumbered,
  pdftitle    = {\titlerunning},
  pdfauthor   = {\authorrunning},
  pdfsubject  = {EPTCS},               
  colorlinks = true,
  linkcolor  = blue!75!black,
  citecolor = blue!75!black
}

\begin{document}
\maketitle

\newcommand{\comp}[1]{{\tt\small\color{orange!75!black} #1}}
\newcommand{\state}[1]{{\tt\small\color{rgb:red,131;green,49;blue,88} #1}} 
\newcommand{\act}[1]{{\tt\small\color{rgb:red,52;green,149;blue,157} #1}} 
\newcommand{\guard}[1]{{\tt\small\color{rgb:red,50;green,158;blue,70} #1}}
\newcommand{\invar}[1]{{\tt\small\color{rgb:red,160;green,56;blue,155} #1}}
\newcommand{\upd}[1]{{\tt\small\color{rgb:red,58;green,56;blue,155} #1}}
\newcommand{\comm}[1]{{\tt\itshape\small\color{rgb:red,171;green,0;blue,48} #1}} 
\newcommand{\stt}[1]{\texttt{{\small #1}}}

\newcommand{\shl}[2][gray]{{\setlength\fboxsep{2pt}\ttfamily\small\colorbox{#1!20}{{#2}}\setlength\fboxsep{2pt}}}
\newcommand{\mshl}[2][gray]{\shl[#1]{\ensuremath{#2}}\xspace}

\section{Introduction}

Model checking real-time systems is complex. This particular work was motivated by and developed in collaboration with an industrial use-case provider: the Alstom railway company, in the context of the VALU3S European project. In this use-case we formally analyse a {motor controller} used in signalling systems: a safety-critical embedded system that reacts to instructions to turn a motor left or right.
Given the criticality of this system and the need to comply to railway standards~\cite{EN50126:2017,EN50128/A2:2020,EN50129:2018}, the motor controller includes redundancy techniques, and its certification requires formal evidences that given time-bounds are met.

The implementation of this motor controller has been developed hand-in-hand with the formal specification of a real-time model in Uppaal~\cite{david2015uppaal}, with a mutual influence between the two. Full details of this use-case can be found in our previous work
~\cite{proenca-verification-2022}. The level of detail and the amount of non-determinism in early models quickly led to state-space explosions when analysing properties such as deadlock freedom. To cope with the state-space explosion problem, different details could be abstracted away.
This led us to two core challenges:
(i) how to efficiently involve both experts in model-checking and experts in the application domain; and
(ii) how to balance trade-offs in the formal specifications between including \emph{enough details} to be faithful to the implementation and \emph{not too many details} to avoid model-checking more complex requirements.

Our approach involves the creation of many variations of the real-time specification, and using MS Excel spreadsheets to help keeping the developers engaged and not interacting directly with the model-checker.
The Uppaal specifications are annotated, and a set of companion spreadsheets controls variability, i.e., for each variation it configures both how the annotated parts of the Uppaal specification can be modified and which requirements should be used.

\myparagraph{Contributions.}
This paper presents extensions that provide a better support for variability, introducing the concept of a feature model~\cite{DBLP:conf/re/SchobbensHT06} within the spreadsheets to validate configurations, and introducing integer attributes to these feature models. We provide a companion open-source tool---\Uppex---that reads MS Excel spreadsheets and Uppaal models and automatises the feature analysis and the model-checking processes. The results are validated within the railway use-case, provided by Alstom, already described in detail in our previous work~\cite{proenca-verification-2022}. We further use a simpler example that the reader can use to experiment with \Uppex.

\myparagraph{Related work}
Model-checking complex systems is difficult and often infeasible due to space explosion. A possible approach to verify properties over networks of automata with a state-space that is too large to traverse is to use statistical model checking (SMC)~\cite{DBLP:series/lncs/LegayLTYSG19}. Uppaal Stratego supports SMC~\cite{david2015uppaal}, and has shown promising results in the railway domain over a moving block signalling system~\cite{basile2022statisticalSTTT}. Using SMC, properties are quantified over the probability of occurring, and model-checking involves performing many runs of the system until the confidence reflects the probability of the property. \Uppex provides an alternative to model-check complex systems, without losing the strength of symbolic model-checking, by facilitating the process of producing many simplifications, each abstracting over different aspects. This family of simpler models is automatically model-checked by successive instantiations and invocations to Uppaal.
Although we use the Uppaal model checker, this tool and our methodology can be easily adapted to other model-checkers such as IMITATOR~\cite{DBLP:conf/cav/Andre21-Long} or mCRL2~\cite{DBLP:conf/fase/BeekVW17}.

The idea of verifying a family of systems efficiently has been investigated and well received in the software product line community~\cite{DBLP:conf/icse/ClassenHSLR10,DBLP:conf/fase/BeekVW17}.
The goal of these approaches is to be able to verify a set of properties in all members of a family of systems. This is often realised by modelling the variability aspect together with the behavioural aspect, avoiding the generation of one model for each member. On the contrary, our approach produces one instance of the model for each member. This creates less dependencies to the choice of the concrete model-checker and allows customising which properties are verified at each instance, at the cost of performance and number of configurations supported. Furthermore, \Uppex attempts to provide an easy interface between modellers and developers, giving the power to developers to fine-tune parameters and configurations without being exposed to the model-checker.

\Uppex uses a Domain Specific Language to represent feature models, for which many textual and modelling languages exist~\cite{DBLP:conf/splc/BeekSE19}. A feature model is here represented as a spreadsheet table, getting inspiration mainly from the UVL language~\cite{DBLP:conf/splc/SundermannFERT21}, but exploiting the tabular representation to capture the tree structure of feature diagrams~\cite{DBLP:conf/re/SchobbensHT06}.

Several approaches exist to realise variability, i.e., to generate software artefacts from a selection of features~\cite{DBLP:journals/infsof/El-SharkawyYS19}.
Popular ones include annotative and compositional approaches~\cite{DBLP:books/daglib/0032924}. Annotative approaches mark code {blocks} that should be removed when some feature is absent at compile time, e.g. using the C-preprocessor to hide {blocks} of code using \stt{\#ifdef} directives. Compositional approaches, such as feature-oriented programming~\cite{DBLP:conf/gttse/Batory06}, aspect-oriented programming~\cite{kiczales1997aspect}, and delta-oriented programming \cite{DBLP:conf/splc/SchaeferBBDT10}, provide mechanisms to inject {blocks} of code based on the selected features.
\Uppex uses annotated {blocks} in a compositional way, i.e., they act as \emph{hooks} marking consecutive lines of the specification file that can be modified when producing variations. This is aligned with the aspect-oriented approach, which uses patterns to discover {blocks} to be adapted (instead of explicit hooks), and with the delta-oriented approach, which uses the names of structural elements (such as classes, objects, and methods) as the {blocks} to be adapted. Our approach is more primitive, in the sense that it is not aware of the structure of the documents being adapted. This makes it more independent of the target language and analyser being used in the back-end, at the cost of understanding and reusing the content of the {blocks} being replaced. For example, \Uppex cannot keep an existing annotated {block} and add a new line, but can only replace the full {block} with a new one.

\myparagraph{Organization of the paper.}
\Cref{sec:motivation} provides more details over our motivating railway scenario prior to our extensions.
\Cref{sec:feat-modelling} describes how to add variability to Uppaal models with \Uppex, using features an feature models, using a simpler example.
\Cref{sec:discussion} summarises some lessons learned when using \Uppex, and
\cref{sec:conclusion} concludes this paper and suggests lines of future~work.

\section{Motivation: model-checking a motor controller}
\label{sec:motivation}

The system under study is a motor controller; its detailed component architecture is depicted in \cref{fig:topology}.
Overall, the controller receives \emph{instructions} from a dashboard (to turn left, to turn right, or a heartbeat), and sends \emph{signals} to a circuit that triggers the corresponding rotation of an engine. The circuit sends periodic \emph{reports} to the controller, either informing that the maximum rotation was reached or that a problem was found. Finally the controller notifies the dashboard whenever an important update or warning exists.

\begin{figure}
  \centering
  \input{src/img/topology}
  \caption{Architecture of the concurrent components being modelled.}
  \label{fig:topology}
\end{figure}

The architecture in \cref{fig:topology} includes other details, explained below.
\begin{itemize}
  \item The system has redundancy: most components are replicated (e.g., \textsf{Controller$_1$} and \textsf{Controller$_2$}, and their consistency is verified by monitors and decoders.
  \item The environment is modelled by 3 components: the \textsf{Dashboard Simulator}, the \textsf{Circuit Simulator}, and the \textsf{Fault Injector}; different scenarios can be considered, to analyse the behaviour under well- and ill-behaved environments.
  \item The components interact in different ways: using synchronisation barriers (\arrow[arrS]), non-blocking synchronous sends that lose data when the reader is not ready (\arrow[arrL]) or that are guaranteed by the receiver to be received (\arrow[arrN]), and asynchronous interaction via a shared variable that is written by the sender and read by the receiver (\arrow[arrD]).
\end{itemize}
The core behaviour is described by both \textsf{Controller} components, who are responsible to detect errors and enter a fallback state in such cases, e.g., when the engines take too long or are too fast to reach the end of a rotation.

Our formal model of this system in Uppaal encodes each component as a state machine, more specifically a real-time automaton~\cite{david2015uppaal}.
When model-checking this model many requirements cannot be verified precisely due to a space explosion. This is because, in many time-points, a very large number of interleavings were possible. 
E.g., often 8 different components could perform some interaction in any possible order. Our solution consists in creating many \emph{variants} of the real-time model, simplifying different aspects of this model, and selecting different requirements to different variants. These variants include, among others:
\begin{itemize}
  \item different environments (dashboards, circuits, and fault injectors);
  \item discarded heartbeat signals, i.e., periodic messages sent from the dashboard to confirm that the motor is available;
  \item discarded consistency checks between replicated counter-parts;
  \item discarded reading from the circuit; and
  \item discarded part of the controller behaviour (initial tests).
\end{itemize}
In total, we collected ±35 aspects that could be toggled, called \emph{features}, and manually selected 14 combinations of these features, called \emph{configurations}. The choice of this combinations was driven by the requirements, i.e., adapted until each requirement could be verified in a rich-enough set of variants. Note that some of these features were describing requirements that must be verified, e.g., if deadlock freedom should be verified. Also note that these features are not meant to be optional features of the implementation, e.g., we always expect the final system to use heartbeats; however, abstracting it away in some variations allows the verification of properties that do not rely on heartbeats.

Statistical Model Checking (SMC), also supported by Uppaal and applied in a similar context~\cite{basile2022statisticalSTTT}, is an alternative approach that we avoid. Using SMC one can verify properties with a given level of certainty, based on many runs of the model. However, it does not provide the same level of certainty of traditional symbolic model-checking.

\subsection*{Automatisation with spreadsheets and \Uppex}

We propose to automatise the verification of these variants,
initially reported in RSSRail 2022~\cite{proenca-verification-2022},
using
(i) \emph{spreadsheets} to represent both core parameters and requirements of the system under study, and
(ii) a prototype tool \Uppex\footnote{{\small\url{https://cister-labs.github.io/uppex/}}} to automatise the creation and verification of variations of the formal specification,
whereas each variation can have a different set of requirements.
Formal models are annotated, specified, and verified using the Uppaal model checker~\cite{david2015uppaal}. Uppaal targets real-time systems, using special variables called clocks that capture the passage of time, and using these clocks to guide the behaviour (with some syntactic restrictions that make the model-checking problem feasible).

\begin{figure}[tb]
  \centering
  \tikz[inner sep=0pt]{
    \tikzstyle{nd}=[line width=2pt,draw=black!70]
    
    \node[nd](a)
        {\includegraphics[width=0.34\textwidth]{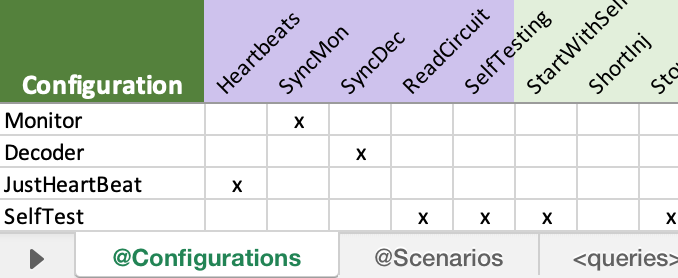}};
    \node[nd,below,yshift=-2pt](b)at(a.south)
        {\includegraphics[width=0.34\textwidth]{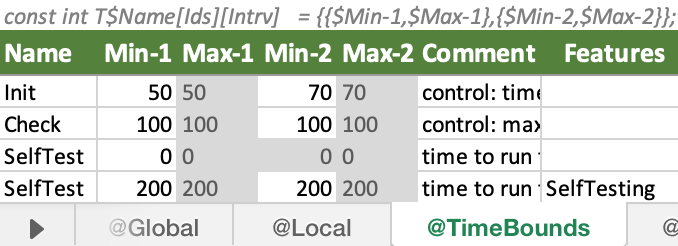}};
    \coordinate(m)at($(a.north)!0.5!(b.south)$);
    \node[nd,right,xshift=2pt](c)at(m-|a.east)
        {\includegraphics[width=0.204\textwidth]{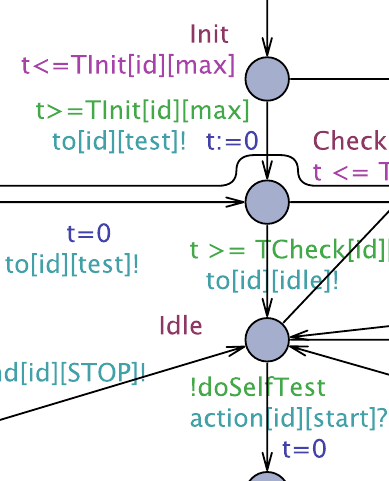}};

    \node[nd,right=3 of c](d)
        {\includegraphics[width=0.194\textwidth]{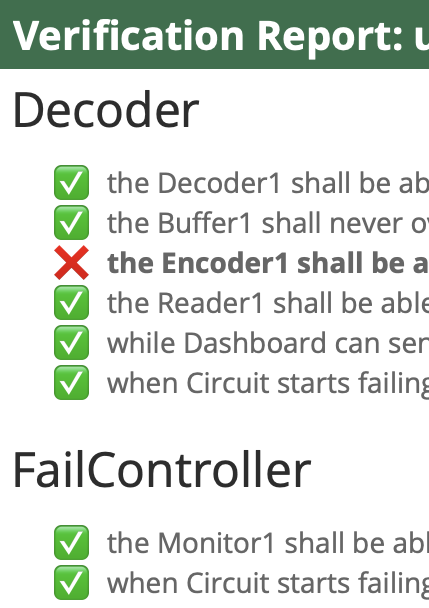}};

    \node(arrow)at($(c)!0.5!(d)$){{\Huge$\Rightarrow$}};
    \node[above,yshift=6pt]at(arrow.north){{\Large{\Uppex}}};
  }
  \caption{\Uppex workflow: updating and verifying models based on configuration tables}
  \label{fig:uppex}
\end{figure}

\Uppex is an open-source command-line tool developed in Scala that
reads both a set of spreadsheets with configurations in MS Excel and an annotated Real-Time specification in Uppaal. Other back-ends are future work, e.g. \textsf{IMITATOR}~\cite{DBLP:conf/cav/Andre21-Long}. A typical workflow is depicted in \cref{fig:uppex}:
given a set of configuring spreadsheets and an annotated Uppaal specification (left), \Uppex produces an \stt{html} report (right) listing properties that passed, failed, or timed-out for each configuration.

More specifically,
\Uppex interprets (1) \emph{special sheets from a MS Excel file} and (2) an \emph{annotated Uppaal file} (XML format), briefly described below. 
\begin{itemize}
  \item Two types of \emph{annotated blocks} are recognised by \Uppex in the Uppaal file: (1) a sequence of consecutive lines starting with ``\comm{// @BlockName}'' until an empty line, such as the ``\comm{// @Limit}'' block on the right of~\cref{fig:hammer-annotated}, and (2) an XML element ``\comm{<BlockName>...</BlockName>}'', covering the text between the tags. Both these annotated blocks have an identifier (the \comm{BlockName}) and a consecutive sequence of lines.

  \item The sheet \stt{@Configurations} (top-left of \cref{fig:uppex}) lists valid combinations of \emph{features}, describing \emph{configuration names} in the first column and \emph{feature names} in the first row.

  \item Any other sheet starting with \stt{@}, such as \stt{@Timebounds} (bottom-left of \cref{fig:uppex}) describe what code will be injected in the Uppaal specification, in this case in an annotated block named \stt{Timebounds}.
  The column named \stt{Features} is used to filter rows based on the selected configuration -- in this case the last two rows have the same identifier (\stt{SelfTest}), and when the \stt{SelfTesting} feature is active the last row will override the previous one. We call these \emph{@-annotations}.

  \item Any sheet with a name surrounded by $\stt{<} \cdot \stt{>}$ (e.g. \stt{<queries>}), is similar to an \stt{@}-sheet, but targetting annotated blocks given by XML elements, as explained above.
  We call these \emph{xml-annotations}.

\end{itemize}

We will describe each of these tables and annotations in more detail below, guided by a simpler example, and extend this approach to further exploit the analyses of features.

\section{Feature modelling in \Uppex}
\label{sec:feat-modelling}

In this work we extend \Uppex to further exploit the feature analysis, introducing \emph{data attributes}, \emph{feature conditions}, and a \emph{feature model}. These are explained below using a simpler but complete example of a \comp{hammer} automaton interacting with a \comp{worker} automata while hitting nails. This example can be found together with the tool at {\small\url{https://github.com/cister-labs/uppex/blob/v0.1.3/examples}}.%

\subsection{Annotating Uppaal specifications}

\begin{figure}
  \centering
  \wrap{\includegraphics[width=0.42\textwidth]{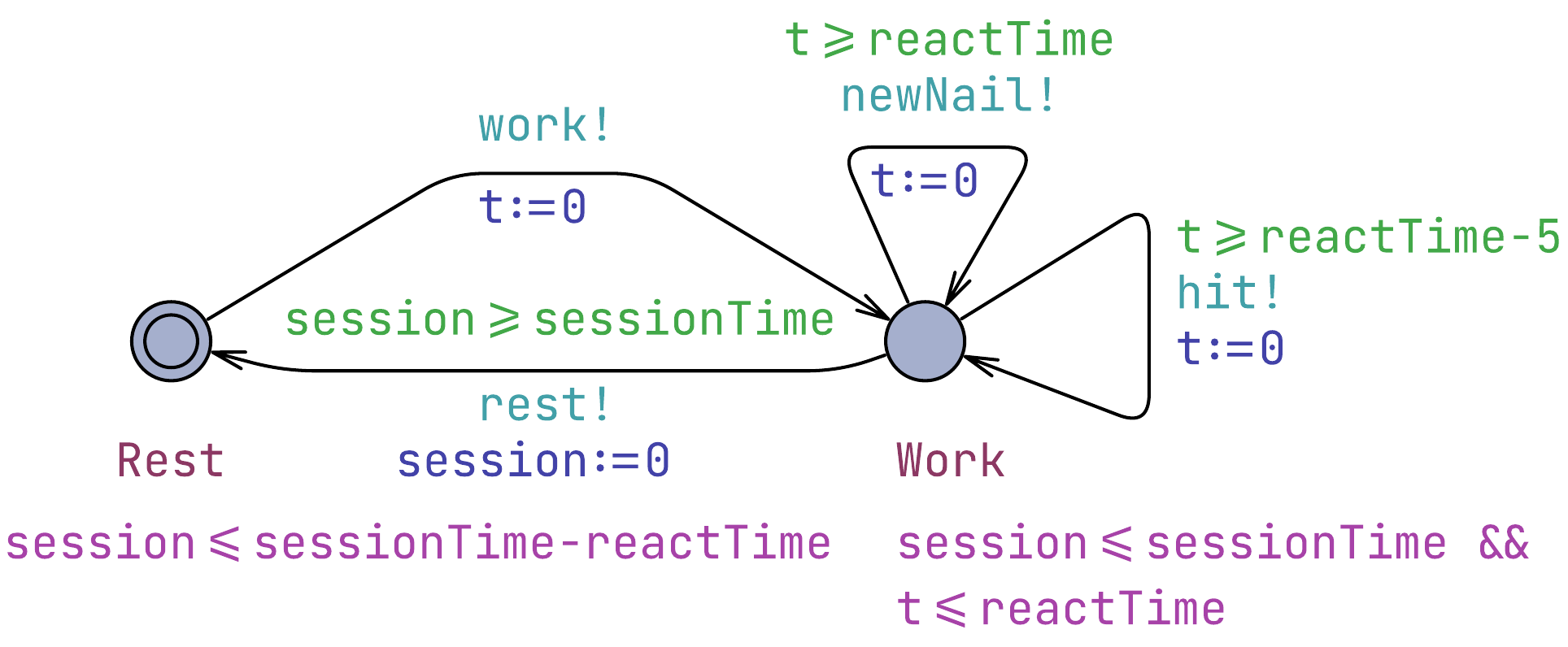}}%
  \wrap{\includegraphics[width=0.27\textwidth]{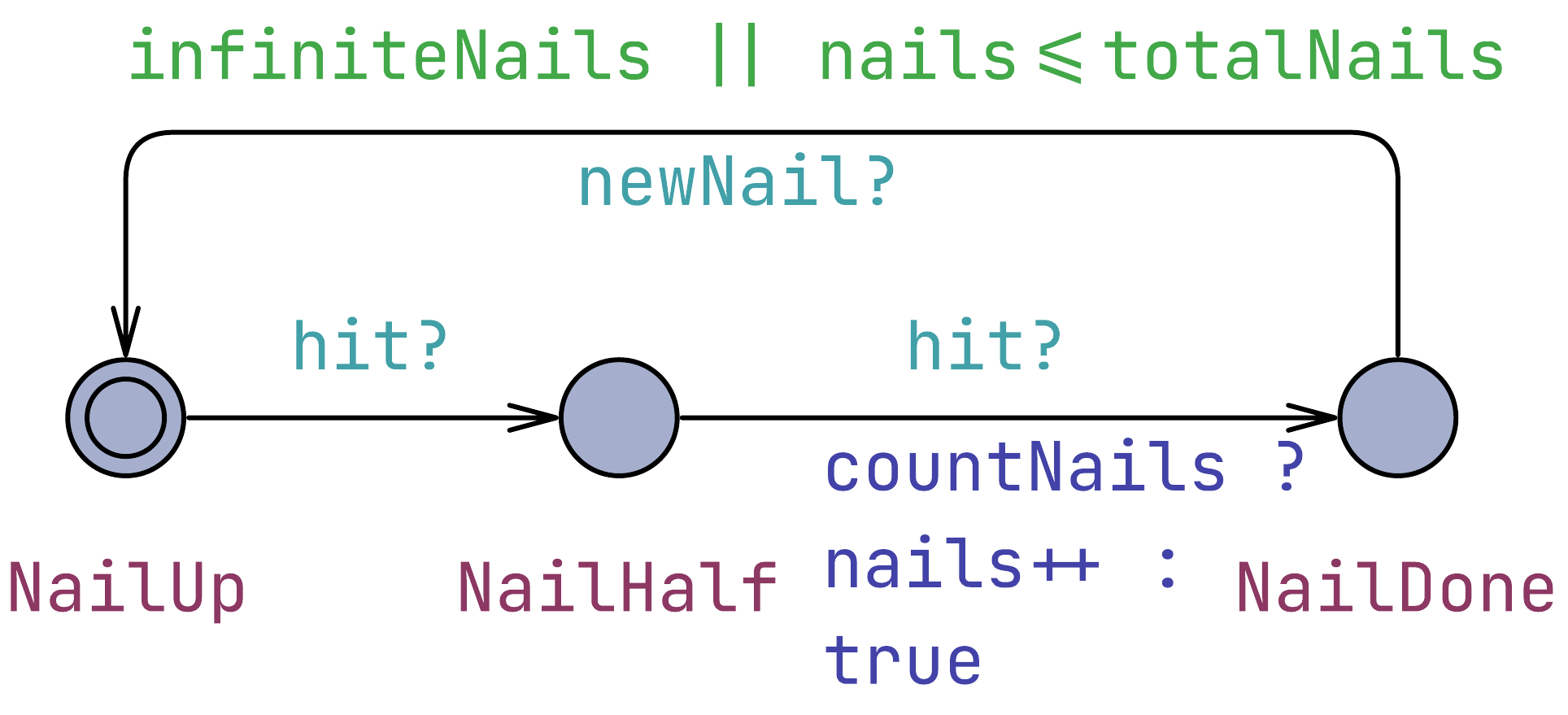}}
  \wrap{\fbox{\includegraphics[width=0.28\textwidth,trim={58.7mm 0mm 0 3mm},clip]{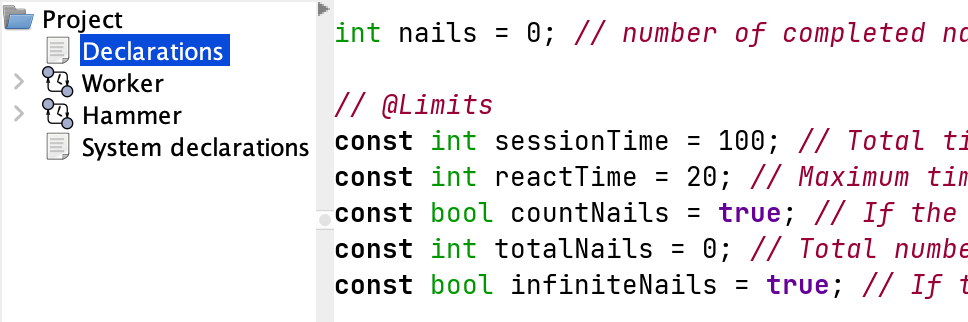}}}
  \caption{Annotated Uppaal specification of a \comp{worker} (left) and a \comp{hammer} (middle); this specification is an XML file with code snippets (c.f. right side) with c-like code that is used by the automata}
  \label{fig:hammer-annotated}
\end{figure}

When developing a family of models with \Uppex, the starting point is a parameterised model. In \cref{fig:hammer-annotated} we present a simple example with 2 timed-automata, where a \comp{worker} is either \state{Rest}ing or \state{Work}ing. While working, it uses a \comp{hammer} to either \act{hit} a nail or to place a \act{newNail}. The code on the right side is used by the Uppaal specification; e.g., \stt{sessionTime} represents the combined time to rest and work by the \comp{worker}, set to 100. The other variables, from top to bottom respectively, capture the maximum time to hit a nail or to add a new one, if the nails should be counted, the number of nails, and if no limit of nails should be considered. The details of the semantics of timed-automata are out of the scope of this paper; intuitively each transition can have a \guard{guard} representing when the transition is active, an \act{action} that will act as a synchronisation barrier with a counterpart action, and an \upd{update} that updates variables after a transition. Some special variables represent time and are called \emph{clocks}; in our example \stt{t} and \stt{session}.

\subsection{Configuring variants}\label{sec:configuring variants}
A \emph{configuration} is a variation of the Uppaal specification by replacing an annotated block 
by a new block with the same name. In our example, the code on the right of \cref{fig:hammer-annotated} has a \stt{@Limits} block with 5 lines. Using a companion MS Excel spreadsheet, we can specify configurations that describe how these annotated blocks can be replaced.

\begin{figure}[hbt!]
  \centering
  \begin{tabular}{@{}|@{~}c@{~}|@{~}c@{~}|@{~}c@{~}|@{}}
  \hline
  ~&~&~\\[-4mm]
  \includegraphics[scale=0.41,trim={0 0 10mm 0},clip]{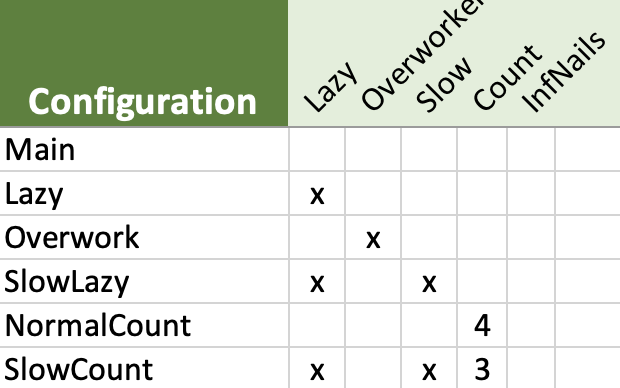}&
  \includegraphics[scale=0.41,trim={0 0 7.8mm 0},clip]{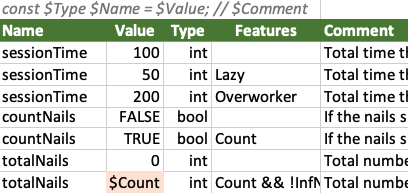}&
  \includegraphics[scale=0.41,trim={0 0 7.8mm 0},clip]{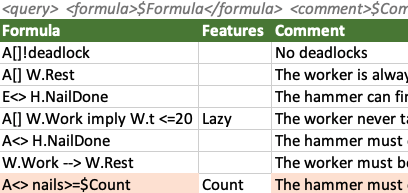}
  \\[-2pt]
  \includegraphics[scale=0.41,trim={0 0 57mm 0},clip]{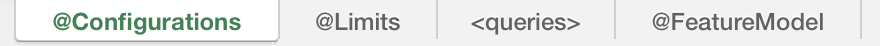}&
  \includegraphics[scale=0.41,trim={0 0 19mm 0},clip]{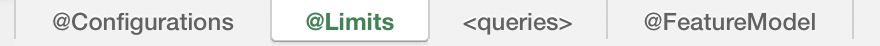}&
  \includegraphics[scale=0.41,trim={0 0 19mm 0},clip]{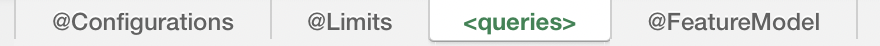}
  \\\hline
  \end{tabular}
  \caption{Defining configurations with spreadsheets: selection of features in \stt{@Configuration} (left), defining the \stt{@Limits} annotation (middle), and defining the \stt{<queries>} annotation (right)
  }
  \label{fig:hammer-sheets}
\end{figure}

The middle of \cref{fig:hammer-sheets} presents the \stt{@Limits} sheet in our hammer example, containing a table of values that is used to produce the associated \stt{@Limits} annotation block.

This table is called an \emph{@-annotation}. In the new {block} each line is formatted according to the top row ``\textcolor{black!50}{\tt\small\itshape const \$Type \$Name = \$Value; // \$Comment}''.
{Blocks} can also refer to XML tags, to replace {blocks} delimited by a given tag; e.g. the sheet on the right of \cref{fig:hammer-sheets} is an \emph{xml-annotation} that specifies a list of requirements using Uppaal's logic
that will replace the content of the \stt{<queries>} XML~element.

A configuration is a set of features, defined in the \stt{@Configuration} table (left of \cref{fig:hammer-sheets}). For example, the configuration named \textsf{SlowLazy} includes the features \textsf{Lazy} and \textsf{Slow}. Features can also have an associated value, e.g., \textsf{Count} is assigned to ``4'' in configuration \textsf{NormalCount} and to ``3'' in \textsf{SlowCount}.

The annotation tables (c.f. middle and right of \cref{fig:hammer-sheets}) can have a special column named \stt{Features} with boolean expressions over feature names. This is used to filter rows: given a configuration, only rows with an expression that holds for the corresponding set of feature is considered. Empty expressions are trivially true. Furthermore, the left-most column acts as an identifier: if more than a row with the same identifier is selected, the last one with a valid feature expression is used.
We chose to use an \emph{overriding} interpretation, instead of forcing these feature expressions to be disjoint for entries with the same identifier, because we found these specifications to be simpler to write and more compact.
In this example selecting the \textsf{Overworker} feature and not \textsf{Lazy} will discard the 2$^{\rm{nd}}$ row for \textsf{sessionTime}, and the 3$^{\rm{rd}}$ will override the 1$^{\rm{st}}$. I.e., the variable \textsf{sessionTime} will be set to 200. The value of a feature can be used in the other cells of a row; e.g., the value of \textsf{totalNails} will be set to 4 when choosing the configuration \textsf{NormalCount}.

This work extends our previous approach~\cite{proenca-verification-2022} by (i) associating values to features and (ii) using of expressions over features instead of individual features in the \textsf{Features} column.

\subsection{Validating features}

\begin{figure}
  \centering
  \begin{tabular}{@{}@{~}c@{~~~~}@{~}c@{~}@{}}
  \wrap{
  \includegraphics[scale=0.41,trim={0 0 0mm 0},clip]{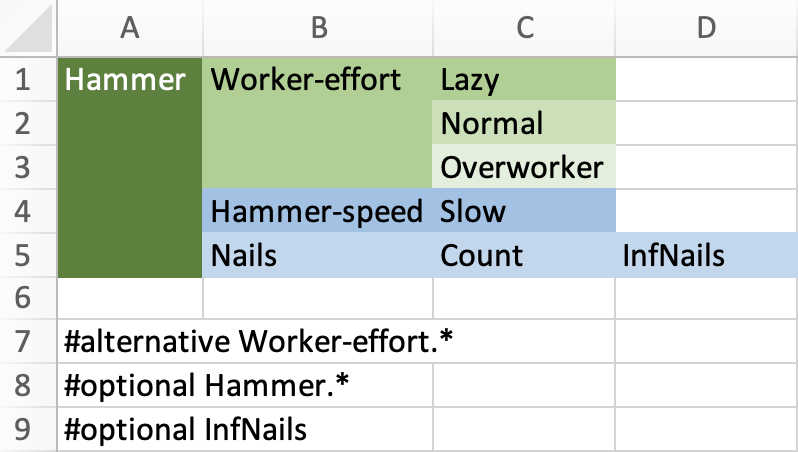}\\
  ~~~~~~\includegraphics[scale=0.41,trim={0 0 0mm 0},clip]{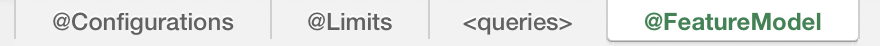}
  }
  &~~~~~
  \wrap{
%
\begin{forest}
  disjunction tree,
  [Hammer, for children={optional,abstract}
    [Worker-effort, alternative={1/3} 
      [Lazy]
      [Normal]
      [Overworker]
    ]
    [Hammer-speed, optional,
      [Slow]
    ]
    [Nails, optional,  
      [Count,  
        [InfNails, optional]
      ]
    ]
  ]
\end{forest}
  }
  \end{tabular}
  \caption{Example of a feature diagram: its tabular form (left) and its usual representation (right)}
  \label{fig:hammer-fm}
\end{figure}

Not all combination of features in the \stt{@Configuration} table should be considered. For example, the worker should not be both lazy and overworker. Such constraints are compiled in another special table called \stt{@FeatureModel}, using a tabular form of feature diagrams~\cite{DBLP:conf/re/SchobbensHT06}. These constraints describe valid combinations of features but are not related to the \stt{Feature} columns in the annotation tables.
We borrowed some constructs from the textual UVL language for feature models~\cite{DBLP:conf/splc/SundermannFERT21}, and synthesise UVL diagrams in \Uppex. An example of a feature diagram can be found in \cref{fig:hammer-fm}: on the left our tabular representation, and on the right its more traditional visual representation.
The table is interpreted as follows.
\begin{itemize}
  \item Non-empty rows whose 1$^{\rm{st}}$ column does not start with \shl{\#} describe the \textbf{tree structure}:
    the parents on the left and the children on the right.
    For example, cell \stt{C4} (\stt{Slow}) is the child of \stt{B4} (\stt{Hammer-speed}), which in turn is the child of \stt{A4}; the latter cell is empty, meaning that it inherits the previous value in column \stt{A}, i.e. \stt{A1} (\stt{Hammer}).

  \item Rows whose 1$^{\rm{st}}$ column starts with \shl{\#} describe a \textbf{constraint}:
  \begin{itemize}
    \item \shl{\#mandatory <siblings>} -- given a set of features with a common parent (\stt{siblings}), it states that these are mandatory whenever the parent is selected;
    \item \shl{\#optional <siblings>} -- states that a set of siblings are optional, even if the parent is selected
    \item \shl{\#alternative <siblings>} -- states that a set of siblings are exclusive and at most one should be included whenever the parent is selected;
    \item \shl{\#or <siblings>} -- states that at least one out of a set of siblings should be included whenever the parent is selected;
    \item \shl{\#constraint <feature-constraint>} -- is a boolean formula over features, following the same syntax as in the \stt{Features} column (c.f. \cref{sec:configuring variants}), that must hold.
  \end{itemize}
\end{itemize}

Only the \stt{\#alternate} and the \stt{\#optional} constraints are illustrated in \cref{fig:hammer-fm}, and by default all features are mandatory. Combining the tree structure and the constraints yields a feature diagram, such as the one on the right of \cref{fig:hammer-fm}. Currently \Uppex supports feature-constraints over features but not over feature attributes, which is left as future work. The tree structure also imposes a strong need to include parent features whenever a child is selected -- \Uppex exploits this by automatically expanding the selection of features to all the parents of the selected ones.

\subsection{Workflow using \Uppex and Uppaal}

So far we described how to specify the input models: (i) the annotated Uppaal specification, (ii) the tables with possible parameters and requirements, (iii) the table with configurations of features, and (iv) the table with the feature model. This subsection describes our proposed \emph{methodology}, i.e., the suggested workflow with \Uppex and Uppaal during the development of a model.

\Uppex's tool is a standalone JAR file \stt{uppex.jar}, open-source and available at \url{https://github.com/cister-labs/uppex/releases}, that can be executed as a command line tool using \shl{java -jar} \shl{uppex.jar [options] <mytables.xlsx>}.
We expect a typical development of a Uppaal+\Uppex project to proceed as follows.
\begin{enumerate}
  \item \textbf{Model:} \label{step:sim}
  Produce a base Uppaal model \stt{project.xml}, i.e., a network of timed automata that can be simulated in Uppaal.
  \\
  \emph{Edit: \underline{Automata in Uppaal}}

  \item \textbf{Parameterise:} \label{step:par}
  Identify a set of parameters that can be useful to expose to domain experts and create the associated @-annotations in the companion Excel file \stt{project.xlsx}; update the Uppaal model by running \Uppex with no arguments, e.g. \shl{java -jar uppex.jar project.xlsx}.
  \\
  \emph{Edit: \underline{Automata in Uppaal} \& \underline{@-annotations in Excel}}

  \item \textbf{Verify behaviour:} \label{step:verbeh}
  Identify a set of requirements, specify them using Uppaal's CTL, and place these in the \stt{<queries>} spreadsheet (c.f. right of \cref{fig:hammer-sheets}); update the Uppaal model as before, or verify all properties using \Uppex using the command \shl{java -jar uppex.jar {-}{-}run project.xlsx}.
  \\
  \emph{Edit: \underline{\stt{<queries>}-annotation in Excel}}

  \item \textbf{Instantiate:}
  Identify variability points and features, populating the annotation tables with a column \stt{Features} (c.f. right of \cref{fig:hammer-sheets}); create the \stt{@Configurations} table to list products, i.e.,  desired combinations of features (c.f. left of \cref{fig:hammer-sheets}); transform the working Uppaal file to~match any given configuration (or product) \shl{prod} by running \shl{java -jar uppex.jar -p prod project.xlsx}; the verification in step (\ref{step:verbeh}) can also receive the \shl{-p prod} option, or simply \shl{{-}-runAll} to verify all available products.
  \\
  \emph{Edit: \underline{\stt{@Configurations}} \& \underline{annotations in Excel}}

  \item \textbf{Verify instances:}
  Identify restrictions over what features can be combined, and specify these in the special \stt{@FeatureModel} table (c.f. \cref{fig:hammer-fm}); \Uppex will always validate all features when verifying or applying a product, but can also be used exclusively for validation by running \shl{java -jar} \shl{{-}{-}validate project.xlsx}.
  \\
  \emph{Edit: \underline{\stt{@FeatureModel}} in Excel}

\end{enumerate}

\begin{figure}[tb!]
  \centering
  \begin{minipage}{0.41\textwidth}
  \begin{lstlisting}
>>> java -jar uppex.jar \
    --runAll hammer.xlsx
features
  Hammer 
    optional
      Worker-effort 
        alternative
          Lazy 
          Normal 
          Overworker 
      Hammer-speed 
        mandatory
          Slow 
      Nails 
        mandatory
          Count 
            optional
              InfNails 
constraints
  !(Lazy && Overworker)
---
 - Products: InfiniteCount, NormalCount, SlowCount, SlowLazy, Lazy, Slow, Overwork, Main
> Reading Uppaal file 'hammer.xml'
---Verifying 'InfiniteCount'---
  | Error or time-out after 30s. Missing 7 properties. Failed on:
  | "No deadlocks"
---Verifying 'NormalCount'---
[FAIL] No deadlocks
[FAIL] The worker is always resting
[OK] The hammer can finish a nail
[FAIL] The hammer must complete a nail
...
  \end{lstlisting}
  \end{minipage}
  ~
  \raisebox{1.5pt}{\wrap{\tikz[inner sep=0pt]{
    \tikzstyle{nd}=[line width=2pt,draw=black!70]
    \node[nd](a)
        {\includegraphics[width=.57\textwidth,trim={0 3pt 0 0},clip]{
          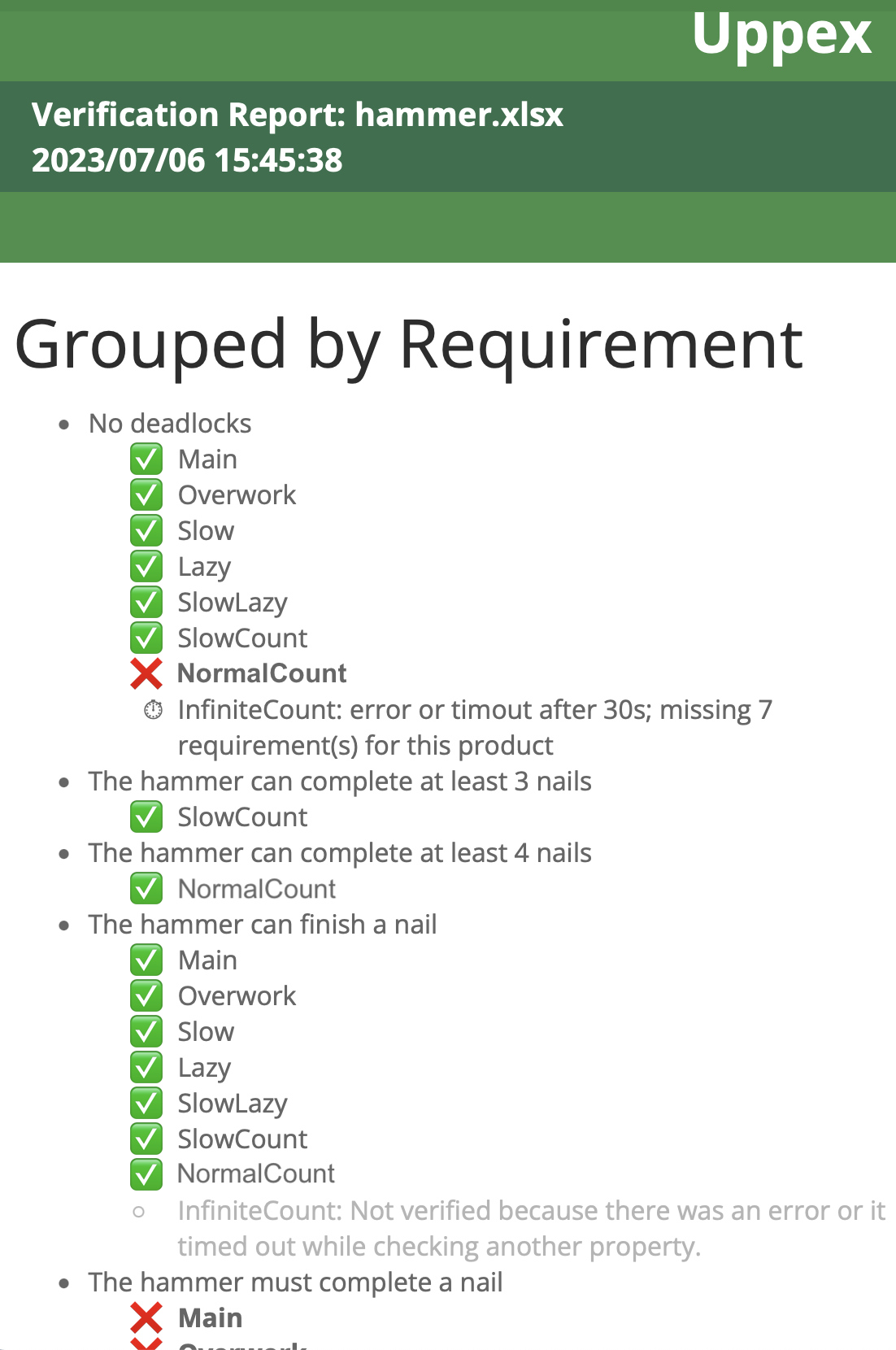}};
  }}}
  \caption{Output when running \Uppex to verify all properties in the hammer project: to the prompt (left) and to the \stt{report.html} file (right)}
  \label{fig:output}
\end{figure}

At each of the steps above it is often needed to revisit the previous steps. E.g., after the verification step~(\ref{step:verbeh}) we expect to be needed to revisit the model in steps~(\ref{step:sim}) and~(\ref{step:par}), to adapt it based on the verification results. 

When a product is applied, a backup of the original version is stored in a folder \stt{backups}, to prevent losing parameters by mistake. This resembles a naïve implementation of a version-control system, where applying a product modifies the working document, while keeping the history of previous versions.

Verifying properties with Uppaal requires the \shl{verifyta} tool to be available at the command line, called by \Uppex using system calls.%
\footnote{Uppaal is a commercial tool, but freely available for academic partners.}
After verifying all properties of all products with \shl{java -jar} \shl{uppex.jar {-}-runnAll project.xlsx}, the tool presents a summary of annotations and configurations found, the feature model in plain text using UVL~\cite{DBLP:conf/splc/SundermannFERT21}, potential errors when validating products, and the results from verifying each property. Each property is marked as passed, failed, or threw an error (e.g., time-out). Furthermore, a \shl{report.html} file is created that clusters these results in a more useful form. The textual output and the HTML report of our hammer example can be found in \cref{fig:output}.

\section{Discussion} \label{sec:discussion}

Our tool and methodology was first applied to our industrial use-case provided by the Alstom railway company on a signalling system, c.f. \cref{sec:motivation}. Many of our design decisions were motivated by weekly discussions between academics and practitioners.
Some of the insights gained by this collaboration using \Uppex are summarised below.
\begin{itemize}
  \item \textbf{Automata size:} The number of automata, locations, and variables easily increased when adding more details, reaching the 16 automata in \cref{fig:topology}.
  The high level of detail was appreciated by Alstom, as well as the use of variability to enable a precise verification of properties without needing to use statistical model checking approaches.

  \item \textbf{Non-determinism:} The high number of non-determinism resulting from allowing several actions to be taken in any order made it more difficult not only to verify, but also to predict the behaviour of the system. Consequently, a new version of this software is being prepared, with a finer scheduling control that reduces this non-determinism to a minimum.

  \item \textbf{Feature model:} The newly added structure to the features in the feature model contributed to a better understanding and insights of what can be modified in the formal model and how.

  \item \textbf{Attributes:} The possibility of using values in the \stt{@Configurations} table facilitated the experimentation with different parameters without having to search through different tables for the values to update.

  \item \textbf{Optimal configurations:} The possibility of enriching \Uppex to support the search for optimal configurations was considered, but this raised concerns regarding a possible increase of the learning curve to use \Uppex. Introducing generic goal functions and cost values could compromise the ease of adoption of \Uppex.

  \item  \textbf{Feature model size:} As the feature model grows, the number of valid variants grows exponentially with the number of features.
  In \Uppex this was not a concern, since it does neither generate all possible variants nor it attempts to find a variant that obeys some condition. Verifying if a single configuration is valid is computationally simple (linear on the size of the feature model). There is a risk of needing to manually add an increasingly large number of configurations to cover a relevant set of combinations, but we did not encounter this problem in our use-cases.
\end{itemize}

When compared the \Uppex version used in our previous work with Alstom~\cite{proenca-verification-2022}, the industrial partners mainly appreciated the possibility of providing numbers in the \stt{@Configuration} table, avoiding the need to navigate through several other sheets. The added structure to the features brought from the feature model also contributed to a better understanding of what the features precisely captured (and resulted in some restructuring of features). Alstom developers were able to edit a shared Excel spreadsheet to adapt some configuration parameters, and were able to understand the generated \stt{html} report, although the execution of \Uppex with the model-checker was mainly carried by the academic partner.
Furthermore, the usage of feature expressions in the \stt{Features} column instead of single features also simplified our model, avoiding some previously added artificial features used to fine-tune the model.

\section{Conclusion and future work}\label{sec:conclusion}

This paper reports on our recent attempt to include feature models represented in our configuring-spreadsheets in an intuitive way for developers, based on feature diagrams with integer attributes, and on how to exploit these for automatic analysis. This work was developed in collaboration with the Alstom railway company, within the VALU3S European project on verification and validation methods and tools.%

Our experience showed that, on the one hand, it is useful to adapt the formal model and requirements by using a set of spreadsheets with key parameters. On the other hand it also highlighted that the pivotal notion of features was not yet fully exploited. This work includes support for feature models with attributes while preserving the simplicity of our spreadsheet-based interface, and keeping an easy-to-use solution that can be adopted by practitioners. 

\medskip

Based on the feedback from Alstom, possible ideas for future work include the following.
\begin{itemize}
  \item \textbf{Coverage:} Currently it is possible to quickly grasp which configurations can validate each of the properties. However, it is hard to provide insights over how complete is this coverage, i.e., how much of the full system is validated for any given property. Counting the number of such configurations is a simple but not fully satisfactory approach. A better approach would be to \emph{quantify the scope} of a configuration, e.g., how many locations can be reached, or which out of a set of reference reachability properties can be proven.

  \item \textbf{Other analysers:} We use Uppaal as our underlying model checker, but \Uppex is general enough to be applied to other static analysis tools with little effort. For example, by using IMITATOR~\cite{DBLP:conf/cav/Andre21-Long} instead we should be able to verify similar properties with a non-commercial tool and search for optimal parameters, and by using mCRL2~\cite{DBLP:conf/tacas/BunteGKLNVWWW19-Long} instead we should be able to support the verification of properties focused on actions rather than states.

  \item \textbf{Deployment configurations:} The same configurations' table could be used to guide the customisation of deployment scripts, or other configuration files that can introduce the variability choices in the concrete software implementations.
\end{itemize}

Furthermore, we invite anyone in the community to submit suggestions or issues using GitHub's issue tracker system, or to contact us for future collaborations.

\subsection*{Acknowledgments}
This work was supported by the CISTER Research Unit (UIDP/UIDB/04234/2020), financed by National Funds through FCT/MCTES (Portuguese Foundation for Science and Technology) and by project IBEX (PTDC/CCI-COM/4280/2021) financed by national funds through FCT. It is also a result of the work developed under projects and Route 25 (ref. TRB/2022/00061 -- C645463824-00000063) funded by the EU/Next Generation, within the Recovery and Resilience Plan (RRP); and project VALU3S (ECSEL/0016/2019 -- JU grant nr. 876852) financed by national funds through FCT and European funds through the EU ECSEL JU. The JU receives support from the European Union’s Horizon 2020 research and innovation programme and Austria, Sweden, Spain, Italy, France, Portugal, Ireland, Finland, Slovenia, Poland, Netherlands, Turkey -- Disclaimer: This document reflects only the author's view and the Commission is not responsible for any use that may be made of the information it contains.

\bibliographystyle{eptcs}  
\bibliography{src/bib}

\begin{thebibliography}{10}
\providecommand{\bibitemdeclare}[2]{}
\providecommand{\surnamestart}{}
\providecommand{\surnameend}{}
\providecommand{\urlprefix}{Available at }
\providecommand{\url}[1]{\texttt{#1}}
\providecommand{\href}[2]{\texttt{#2}}
\providecommand{\urlalt}[2]{\href{#1}{#2}}
\providecommand{\doi}[1]{doi:\urlalt{https://doi.org/#1}{#1}}
\providecommand{\eprint}[1]{arXiv:\urlalt{https://arxiv.org/abs/#1}{#1}}
\providecommand{\bibinfo}[2]{#2}

\bibitemdeclare{inproceedings}{DBLP:conf/cav/Andre21-Long}
\bibitem{DBLP:conf/cav/Andre21-Long}
\bibinfo{author}{{\'{E}}tienne \surnamestart Andr{\'{e}}\surnameend}
  (\bibinfo{year}{2021}): \emph{\bibinfo{title}{{IMITATOR} 3: Synthesis of
  Timing Parameters Beyond Decidability}}.
\newblock In \bibinfo{editor}{Alexandra \surnamestart Silva\surnameend} \&
  \bibinfo{editor}{K.~Rustan~M. \surnamestart Leino\surnameend}, editors:
  {\slshape \bibinfo{booktitle}{Computer Aided Verification - 33rd
  International Conference, {CAV} 2021, Virtual Event, July 20-23, 2021,
  Proceedings, Part {I}}}, {\slshape \bibinfo{series}{LNCS}}
  \bibinfo{volume}{12759}, \bibinfo{publisher}{Springer}, pp.
  \bibinfo{pages}{552--565}, \doi{10.1007/978-3-030-81685-8\_26}.

\bibitemdeclare{book}{DBLP:books/daglib/0032924}
\bibitem{DBLP:books/daglib/0032924}
\bibinfo{author}{Sven \surnamestart Apel\surnameend}, \bibinfo{author}{Don~S.
  \surnamestart Batory\surnameend}, \bibinfo{author}{Christian \surnamestart
  K{\"{a}}stner\surnameend} \& \bibinfo{author}{Gunter \surnamestart
  Saake\surnameend} (\bibinfo{year}{2013}):
  \emph{\bibinfo{title}{Feature-Oriented Software Product Lines - Concepts and
  Implementation}}.
\newblock \bibinfo{publisher}{Springer}, \doi{10.1007/978-3-642-37521-7}.

\bibitemdeclare{article}{basile2022statisticalSTTT}
\bibitem{basile2022statisticalSTTT}
\bibinfo{author}{Davide \surnamestart Basile\surnameend},
  \bibinfo{author}{Maurice~H. \surnamestart ter Beek\surnameend},
  \bibinfo{author}{Alessio \surnamestart Ferrari\surnameend} \&
  \bibinfo{author}{Axel \surnamestart Legay\surnameend} (\bibinfo{year}{2022}):
  \emph{\bibinfo{title}{Exploring the {ERTMS/ETCS} full moving block
  specification: an experience with formal methods}}.
\newblock {\slshape \bibinfo{journal}{Int. J. Softw. Tools Technol. Transf.}}
  \bibinfo{volume}{24}(\bibinfo{number}{3}), pp. \bibinfo{pages}{351--370},
  \doi{10.1007/s10009-022-00653-3}.

\bibitemdeclare{inproceedings}{DBLP:conf/gttse/Batory06}
\bibitem{DBLP:conf/gttse/Batory06}
\bibinfo{author}{Don~S. \surnamestart Batory\surnameend}
  (\bibinfo{year}{2005}): \emph{\bibinfo{title}{A Tutorial on Feature Oriented
  Programming and the {AHEAD} Tool Suite}}.
\newblock In \bibinfo{editor}{Ralf \surnamestart L{\"{a}}mmel\surnameend},
  \bibinfo{editor}{Jo{\~{a}}o \surnamestart Saraiva\surnameend} \&
  \bibinfo{editor}{Joost \surnamestart Visser\surnameend}, editors: {\slshape
  \bibinfo{booktitle}{Generative and Transformational Techniques in Software
  Engineering, International Summer School, {GTTSE} 2005, Braga, Portugal, July
  4-8, 2005. Revised Papers}}, {\slshape \bibinfo{series}{Lecture Notes in
  Computer Science}} \bibinfo{volume}{4143}, \bibinfo{publisher}{Springer}, pp.
  \bibinfo{pages}{3--35}, \doi{10.1007/11877028\_1}.

\bibitemdeclare{inproceedings}{DBLP:conf/splc/BeekSE19}
\bibitem{DBLP:conf/splc/BeekSE19}
\bibinfo{author}{Maurice~H. \surnamestart ter Beek\surnameend},
  \bibinfo{author}{Klaus \surnamestart Schmid\surnameend} \&
  \bibinfo{author}{Holger \surnamestart Eichelberger\surnameend}
  (\bibinfo{year}{2019}): \emph{\bibinfo{title}{Textual variability modeling
  languages: an overview and considerations}}.
\newblock In \bibinfo{editor}{Carlos \surnamestart Cetina\surnameend},
  \bibinfo{editor}{Oscar \surnamestart D{\'{\i}}az\surnameend},
  \bibinfo{editor}{Laurence \surnamestart Duchien\surnameend},
  \bibinfo{editor}{Marianne \surnamestart Huchard\surnameend},
  \bibinfo{editor}{Rick \surnamestart Rabiser\surnameend},
  \bibinfo{editor}{Camille \surnamestart Salinesi\surnameend},
  \bibinfo{editor}{Christoph \surnamestart Seidl\surnameend},
  \bibinfo{editor}{Xhevahire \surnamestart T{\"{e}}rnava\surnameend},
  \bibinfo{editor}{Leopoldo \surnamestart Teixeira\surnameend},
  \bibinfo{editor}{Thomas \surnamestart Th{\"{u}}m\surnameend} \&
  \bibinfo{editor}{Tewfik \surnamestart Ziadi\surnameend}, editors: {\slshape
  \bibinfo{booktitle}{Proceedings of the 23rd International Systems and
  Software Product Line Conference, {SPLC} 2019, Volume B, Paris, France,
  September 9-13, 2019}}, \bibinfo{publisher}{{ACM}}, pp.
  \bibinfo{pages}{82:1--82:7}, \doi{10.1145/3307630.3342398}.

\bibitemdeclare{inproceedings}{DBLP:conf/fase/BeekVW17}
\bibitem{DBLP:conf/fase/BeekVW17}
\bibinfo{author}{Maurice~H. \surnamestart ter Beek\surnameend},
  \bibinfo{author}{Erik~P. \surnamestart de~Vink\surnameend} \&
  \bibinfo{author}{Tim A.~C. \surnamestart Willemse\surnameend}
  (\bibinfo{year}{2017}): \emph{\bibinfo{title}{Family-Based Model Checking
  with mCRL2}}.
\newblock In \bibinfo{editor}{Marieke \surnamestart Huisman\surnameend} \&
  \bibinfo{editor}{Julia \surnamestart Rubin\surnameend}, editors: {\slshape
  \bibinfo{booktitle}{Fundamental Approaches to Software Engineering - 20th
  International Conference, {FASE} 2017, Held as Part of the European Joint
  Conferences on Theory and Practice of Software, {ETAPS} 2017, Uppsala,
  Sweden, April 22-29, 2017, Proceedings}}, {\slshape \bibinfo{series}{Lecture
  Notes in Computer Science}} \bibinfo{volume}{10202},
  \bibinfo{publisher}{Springer}, pp. \bibinfo{pages}{387--405},
  \doi{10.1007/978-3-662-54494-5\_23}.

\bibitemdeclare{inproceedings}{DBLP:conf/tacas/BunteGKLNVWWW19-Long}
\bibitem{DBLP:conf/tacas/BunteGKLNVWWW19-Long}
\bibinfo{author}{Olav \surnamestart Bunte\surnameend},
  \bibinfo{author}{Jan~Friso \surnamestart Groote\surnameend},
  \bibinfo{author}{Jeroen J.~A. \surnamestart Keiren\surnameend},
  \bibinfo{author}{Maurice \surnamestart Laveaux\surnameend},
  \bibinfo{author}{Thomas \surnamestart Neele\surnameend},
  \bibinfo{author}{Erik~P. \surnamestart de~Vink\surnameend},
  \bibinfo{author}{Wieger \surnamestart Wesselink\surnameend},
  \bibinfo{author}{Anton \surnamestart Wijs\surnameend} \& \bibinfo{author}{Tim
  A.~C. \surnamestart Willemse\surnameend} (\bibinfo{year}{2019}):
  \emph{\bibinfo{title}{The mCRL2 Toolset for Analysing Concurrent Systems -
  Improvements in Expressivity and Usability}}.
\newblock In \bibinfo{editor}{Tom{\'{a}}s \surnamestart Vojnar\surnameend} \&
  \bibinfo{editor}{Lijun \surnamestart Zhang\surnameend}, editors: {\slshape
  \bibinfo{booktitle}{Tools and Algorithms for the Construction and Analysis of
  Systems - 25th International Conference, {TACAS} 2019, Held as Part of the
  European Joint Conferences on Theory and Practice of Software, {ETAPS} 2019,
  Prague, Czech Republic, April 6-11, 2019, Proceedings, Part {II}}}, {\slshape
  \bibinfo{series}{LNCS}} \bibinfo{volume}{11428},
  \bibinfo{publisher}{Springer}, pp. \bibinfo{pages}{21--39},
  \doi{10.1007/978-3-030-17465-1\_2}.

\bibitemdeclare{inproceedings}{DBLP:conf/icse/ClassenHSLR10}
\bibitem{DBLP:conf/icse/ClassenHSLR10}
\bibinfo{author}{Andreas \surnamestart Classen\surnameend},
  \bibinfo{author}{Patrick \surnamestart Heymans\surnameend},
  \bibinfo{author}{Pierre{-}Yves \surnamestart Schobbens\surnameend},
  \bibinfo{author}{Axel \surnamestart Legay\surnameend} \&
  \bibinfo{author}{Jean{-}Fran{\c{c}}ois \surnamestart Raskin\surnameend}
  (\bibinfo{year}{2010}): \emph{\bibinfo{title}{Model checking lots of systems:
  efficient verification of temporal properties in software product lines}}.
\newblock In \bibinfo{editor}{Jeff \surnamestart Kramer\surnameend},
  \bibinfo{editor}{Judith \surnamestart Bishop\surnameend},
  \bibinfo{editor}{Premkumar~T. \surnamestart Devanbu\surnameend} \&
  \bibinfo{editor}{Sebasti{\'{a}}n \surnamestart Uchitel\surnameend}, editors:
  {\slshape \bibinfo{booktitle}{Proceedings of the 32nd {ACM/IEEE}
  International Conference on Software Engineering - Volume 1, {ICSE} 2010,
  Cape Town, South Africa, 1-8 May 2010}}, \bibinfo{publisher}{{ACM}}, pp.
  \bibinfo{pages}{335--344}, \doi{10.1145/1806799.1806850}.

\bibitemdeclare{article}{david2015uppaal}
\bibitem{david2015uppaal}
\bibinfo{author}{Alexandre \surnamestart David\surnameend},
  \bibinfo{author}{Kim~G \surnamestart Larsen\surnameend},
  \bibinfo{author}{Axel \surnamestart Legay\surnameend},
  \bibinfo{author}{Marius \surnamestart Miku{\v{c}}ionis\surnameend} \&
  \bibinfo{author}{Danny~B{\o}gsted \surnamestart Poulsen\surnameend}
  (\bibinfo{year}{2015}): \emph{\bibinfo{title}{Uppaal SMC tutorial}}.
\newblock {\slshape \bibinfo{journal}{International journal on software tools
  for technology transfer}} \bibinfo{volume}{17}, pp.
  \bibinfo{pages}{397--415}, \doi{10.1007/s10009-014-0361-y}.

\bibitemdeclare{article}{DBLP:journals/infsof/El-SharkawyYS19}
\bibitem{DBLP:journals/infsof/El-SharkawyYS19}
\bibinfo{author}{Sascha \surnamestart El{-}Sharkawy\surnameend},
  \bibinfo{author}{Nozomi \surnamestart Yamagishi{-}Eichler\surnameend} \&
  \bibinfo{author}{Klaus \surnamestart Schmid\surnameend}
  (\bibinfo{year}{2019}): \emph{\bibinfo{title}{Metrics for analyzing
  variability and its implementation in software product lines: {A} systematic
  literature review}}.
\newblock {\slshape \bibinfo{journal}{Inf. Softw. Technol.}}
  \bibinfo{volume}{106}, pp. \bibinfo{pages}{1--30},
  \doi{10.1016/j.infsof.2018.08.015}.

\bibitemdeclare{techreport}{EN50126:2017}
\bibitem{EN50126:2017}
 (\bibinfo{year}{2017}): \emph{\bibinfo{title}{{Railway Applications. The
  Specification and Demonstration of Reliability, Availability, Maintainability
  and Safety (RAMS). Generic RAMS Process}}}.
\newblock \bibinfo{type}{Standard ({N})}, \bibinfo{institution}{CENELEC}.

\bibitemdeclare{techreport}{EN50128/A2:2020}
\bibitem{EN50128/A2:2020}
 (\bibinfo{year}{2020}): \emph{\bibinfo{title}{{Railway applications.
  Communication, signalling and processing systems - Software for railway
  control and protection systems}}}.
\newblock \bibinfo{type}{Standard ({N})}, \bibinfo{institution}{CENELEC}.

\bibitemdeclare{techreport}{EN50129:2018}
\bibitem{EN50129:2018}
 (\bibinfo{year}{2018}): \emph{\bibinfo{title}{{Railway applications.
  Communication, signalling and processing systems. Safety related electronic
  systems for signalling}}}.
\newblock \bibinfo{type}{Standard ({N})}, \bibinfo{institution}{CENELEC}.

\bibitemdeclare{inproceedings}{kiczales1997aspect}
\bibitem{kiczales1997aspect}
\bibinfo{author}{Gregor \surnamestart Kiczales\surnameend},
  \bibinfo{author}{John \surnamestart Lamping\surnameend},
  \bibinfo{author}{Anurag \surnamestart Mendhekar\surnameend},
  \bibinfo{author}{Chris \surnamestart Maeda\surnameend},
  \bibinfo{author}{Cristina \surnamestart Lopes\surnameend},
  \bibinfo{author}{Jean-Marc \surnamestart Loingtier\surnameend} \&
  \bibinfo{author}{John \surnamestart Irwin\surnameend} (\bibinfo{year}{1997}):
  \emph{\bibinfo{title}{Aspect-oriented programming}}.
\newblock In: {\slshape \bibinfo{booktitle}{ECOOP'97—Object-Oriented
  Programming: 11th European Conference Jyv{\"a}skyl{\"a}, Finland, June 9--13,
  1997 Proceedings 11}}, \bibinfo{organization}{Springer}, pp.
  \bibinfo{pages}{220--242}, \doi{10.1007/BFb0053381}.

\bibitemdeclare{incollection}{DBLP:series/lncs/LegayLTYSG19}
\bibitem{DBLP:series/lncs/LegayLTYSG19}
\bibinfo{author}{Axel \surnamestart Legay\surnameend}, \bibinfo{author}{Anna
  \surnamestart Lukina\surnameend}, \bibinfo{author}{Louis{-}Marie
  \surnamestart Traonouez\surnameend}, \bibinfo{author}{Junxing \surnamestart
  Yang\surnameend}, \bibinfo{author}{Scott~A. \surnamestart Smolka\surnameend}
  \& \bibinfo{author}{Radu \surnamestart Grosu\surnameend}
  (\bibinfo{year}{2019}): \emph{\bibinfo{title}{Statistical Model Checking}}.
\newblock In \bibinfo{editor}{Bernhard \surnamestart Steffen\surnameend} \&
  \bibinfo{editor}{Gerhard~J. \surnamestart Woeginger\surnameend}, editors:
  {\slshape \bibinfo{booktitle}{Computing and Software Science - State of the
  Art and Perspectives}}, {\slshape \bibinfo{series}{Lecture Notes in Computer
  Science}} \bibinfo{volume}{10000}, \bibinfo{publisher}{Springer}, pp.
  \bibinfo{pages}{478--504}, \doi{10.1007/978-3-319-91908-9\_23}.

\bibitemdeclare{inproceedings}{proenca-verification-2022}
\bibitem{proenca-verification-2022}
\bibinfo{author}{José \surnamestart Proença\surnameend},
  \bibinfo{author}{Sina \surnamestart Borrami\surnameend},
  \bibinfo{author}{Jorge~Sanchez \surnamestart de~Nova\surnameend},
  \bibinfo{author}{David \surnamestart Pereira\surnameend} \&
  \bibinfo{author}{Giann~Spilere \surnamestart Nandi\surnameend}
  (\bibinfo{year}{2022}): \emph{\bibinfo{title}{Verification of Multiple Models
  of a Safety-Critical Motor Controller in Railway Systems}}.
\newblock In \bibinfo{editor}{Simon~Collart \surnamestart
  Dutilleul\surnameend}, \bibinfo{editor}{Anne~E. \surnamestart
  Haxthausen\surnameend} \& \bibinfo{editor}{Thierry \surnamestart
  Lecomte\surnameend}, editors: {\slshape \bibinfo{booktitle}{Reliability,
  Safety, and Security of Railway Systems. Modelling, Analysis, Verification,
  and Certification - 4th International Conference, RSSRail 2022, Paris,
  France, June 1-2, 2022, Proceedings}}, {\slshape \bibinfo{series}{Lecture
  Notes in Computer Science}} \bibinfo{volume}{13294},
  \bibinfo{publisher}{Springer}, pp. \bibinfo{pages}{83--94},
  \doi{10.1007/978-3-031-05814-1_6}.

\bibitemdeclare{inproceedings}{DBLP:conf/splc/SchaeferBBDT10}
\bibitem{DBLP:conf/splc/SchaeferBBDT10}
\bibinfo{author}{Ina \surnamestart Schaefer\surnameend},
  \bibinfo{author}{Lorenzo \surnamestart Bettini\surnameend},
  \bibinfo{author}{Viviana \surnamestart Bono\surnameend},
  \bibinfo{author}{Ferruccio \surnamestart Damiani\surnameend} \&
  \bibinfo{author}{Nico \surnamestart Tanzarella\surnameend}
  (\bibinfo{year}{2010}): \emph{\bibinfo{title}{Delta-Oriented Programming of
  Software Product Lines}}.
\newblock In \bibinfo{editor}{Jan \surnamestart Bosch\surnameend} \&
  \bibinfo{editor}{Jaejoon \surnamestart Lee\surnameend}, editors: {\slshape
  \bibinfo{booktitle}{Software Product Lines: Going Beyond - 14th International
  Conference, {SPLC} 2010, Jeju Island, South Korea, September 13-17, 2010.
  Proceedings}}, {\slshape \bibinfo{series}{Lecture Notes in Computer Science}}
  \bibinfo{volume}{6287}, \bibinfo{publisher}{Springer}, pp.
  \bibinfo{pages}{77--91}, \doi{10.1007/978-3-642-15579-6\_6}.

\bibitemdeclare{inproceedings}{DBLP:conf/re/SchobbensHT06}
\bibitem{DBLP:conf/re/SchobbensHT06}
\bibinfo{author}{Pierre{-}Yves \surnamestart Schobbens\surnameend},
  \bibinfo{author}{Patrick \surnamestart Heymans\surnameend} \&
  \bibinfo{author}{Jean{-}Christophe \surnamestart Trigaux\surnameend}
  (\bibinfo{year}{2006}): \emph{\bibinfo{title}{Feature Diagrams: {A} Survey
  and a Formal Semantics}}.
\newblock In: {\slshape \bibinfo{booktitle}{14th {IEEE} International
  Conference on Requirements Engineering {(RE} 2006), 11-15 September 2006,
  Minneapolis/St.Paul, Minnesota, {USA}}}, \bibinfo{publisher}{{IEEE} Computer
  Society}, pp. \bibinfo{pages}{136--145}, \doi{10.1109/RE.2006.23}.

\bibitemdeclare{inproceedings}{DBLP:conf/splc/SundermannFERT21}
\bibitem{DBLP:conf/splc/SundermannFERT21}
\bibinfo{author}{Chico \surnamestart Sundermann\surnameend},
  \bibinfo{author}{Kevin \surnamestart Feichtinger\surnameend},
  \bibinfo{author}{Dominik \surnamestart Engelhardt\surnameend},
  \bibinfo{author}{Rick \surnamestart Rabiser\surnameend} \&
  \bibinfo{author}{Thomas \surnamestart Th{\"{u}}m\surnameend}
  (\bibinfo{year}{2021}): \emph{\bibinfo{title}{Yet another textual variability
  language?: a community effort towards a unified language}}.
\newblock In \bibinfo{editor}{Mohammad~Reza \surnamestart Mousavi\surnameend}
  \& \bibinfo{editor}{Pierre{-}Yves \surnamestart Schobbens\surnameend},
  editors: {\slshape \bibinfo{booktitle}{{SPLC} '21: 25th {ACM} International
  Systems and Software Product Line Conference, Leicester, United Kingdom,
  September 6-11, 2021, Volume {A}}}, \bibinfo{publisher}{{ACM}}, pp.
  \bibinfo{pages}{136--147}, \doi{10.1145/3461001.3471145}.

\end{thebibliography}

\end{document}